\documentclass[11pt]{article}
\usepackage{latexsym}

\def\AFOUR{%
\setlength{\textheight}{8.5in}%
\setlength{\textwidth}{5.75in}%
\setlength{\topmargin}{-0.375in}%
\hoffset=-.5in%
\renewcommand{\baselinestretch}{1.17}%
\setlength{\parskip}{6pt plus 2pt}%
}

\AFOUR                                           

\expandafter\ifx\csname amssym.def\endcsname\relax \else\endinput\fi
\expandafter\edef\csname amssym.def\endcsname{%
       \catcode`\noexpand\@=\the\catcode`\@\space}
\catcode`\@=11
\def\undefine#1{\let#1\undefined}
\def\newsymbol#1#2#3#4#5{\let\next@\relax
 \ifnum#2=\@ne\let\next@\msafam@\else
 \ifnum#2=\tw@\let\next@\msbfam@\fi\fi
 \mathchardef#1="#3\next@#4#5}
\def\mathhexbox@#1#2#3{\relax
 \ifmmode\mathpalette{}{\m@th\mathchar"#1#2#3}%
 \else\leavevmode\hbox{$\m@th\mathchar"#1#2#3$}\fi}
\def\hexnumber@#1{\ifcase#1 0\or 1\or 2\or 3\or 4\or 5\or 6\or 7\or 8\or
 9\or A\or B\or C\or D\or E\or F\fi}

\font\tenmsa=msam10
\font\sevenmsa=msam7
\font\fivemsa=msam5
\newfam\msafam
\textfont\msafam=\tenmsa
\scriptfont\msafam=\sevenmsa
\scriptscriptfont\msafam=\fivemsa
\edef\msafam@{\hexnumber@\msafam}
\mathchardef\dabar@"0\msafam@39
\def\dashrightarrow{\mathrel{\dabar@\dabar@\mathchar"0\msafam@4B}}
\def\dashleftarrow{\mathrel{\mathchar"0\msafam@4C\dabar@\dabar@}}

\def\ulcorner{\delimiter"4\msafam@70\msafam@70 }
\def\urcorner{\delimiter"5\msafam@71\msafam@71 }
\def\llcorner{\delimiter"4\msafam@78\msafam@78 }
\def\lrcorner{\delimiter"5\msafam@79\msafam@79 }
\def\yen{{\mathhexbox@\msafam@55}}
\def\checkmark{{\mathhexbox@\msafam@58}}
\def\circledR{{\mathhexbox@\msafam@72}}
\def\maltese{{\mathhexbox@\msafam@7A}}
\def\circledS{{\mathhexbox@\msafam@73}}

\font\tenmsb=msbm10
\font\sevenmsb=msbm7
\font\fivemsb=msbm5
\newfam\msbfam
\textfont\msbfam=\tenmsb
\scriptfont\msbfam=\sevenmsb
\scriptscriptfont\msbfam=\fivemsb
\edef\msbfam@{\hexnumber@\msbfam}
\def\Bbb#1{{\fam\msbfam\relax#1}}
\def\widehat#1{\setbox\z@\hbox{$\m@th#1$}%
 \ifdim\wd\z@>\tw@ em\mathaccent"0\msbfam@5B{#1}%
 \else\mathaccent"0362{#1}\fi}
\def\widetilde#1{\setbox\z@\hbox{$\m@th#1$}%
 \ifdim\wd\z@>\tw@ em\mathaccent"0\msbfam@5D{#1}%
 \else\mathaccent"0365{#1}\fi}
\font\teneufm=eufm10
\font\seveneufm=eufm7
\font\fiveeufm=eufm5
\newfam\eufmfam
\textfont\eufmfam=\teneufm
\scriptfont\eufmfam=\seveneufm
\scriptscriptfont\eufmfam=\fiveeufm
\def\frak#1{{\fam\eufmfam\relax#1}}

\csname amssym.def\endcsname

\makeatletter
\def\section{\@startsection {section}{1}{\z@}{-3.5ex plus -1ex minus
 -.2ex}{2.3ex plus .2ex}{\large\sc}}
\def\subsection{\@startsection{subsection}{2}{\z@}{-3.25ex plus -1ex minus
 -.2ex}{1.5ex plus .2ex}{\normalsize\sc}}
\makeatother

\makeatletter
\@addtoreset{equation}{section}

\makeatother

\newcommand{\nc}{\newcommand}
\newcommand{\rnc}{\renewcommand}

\nc{\subs}[1]{{\vspace*{0.5cm}}%
{\noindent\underline{\small\sc #1}}{\addcontentsline{toc}{subsubsection}{#1}}%
{\vspace*{0.3cm}}}

\nc{\subss}[1]{{\vspace*{0.5cm}}%
{\noindent\underline{\small\sc #1}}%
{\vspace*{0.3cm}}}

\nc{\chap}[1]{{\clearpage}%
\begin{center}%
{\noindent\underline{\large\sc #1}}{\addcontentsline{toc}{section}{#1}}%
\end{center}%
{\vspace*{0.3cm}}}

\nc{\be}{\begin{equation}}
\nc{\ee}{\end{equation}}
\nc{\bea}{\begin{eqnarray}}
\nc{\eea}{\end{eqnarray}}

\nc{\trac}[2]{{\textstyle\frac{#1}{#2}}}

\nc{\ex}[1]{\mbox{e}^{\,\textstyle#1}}

\nc{\CC}{\Bbb{C}}
\nc{\HH}{\Bbb{H}}
\nc{\PP}{\Bbb{P}}
\nc{\RR}{\Bbb{R}}
\nc{\ZZ}{\Bbb{Z}}
\nc{\II}{\Bbb{I}}
\nc{\EE}{\Bbb{E}}

\rnc{\a}{\alpha}
\rnc{\b}{\beta}
\rnc{\d}{\delta}
\nc{\ga}{\gamma}
\nc{\la}{\lambda}
\nc{\f}{\phi}
\nc{\p}{\psi}
\nc{\e}{\eta}
\rnc{\c}{\chi}
\nc{\eps}{\epsilon}
\nc{\om}{\omega}
\nc{\Om}{\Omega}

\nc{\symx}{\circledS}
\newsymbol\smallsmile 1360
\newsymbol\smallfrown 1361
\nc{\ad}{\mathop{\mbox{ad}}\nolimits}
\nc{\tr}{\mathop{\mbox{tr}}\nolimits}
\nc{\Tr}{\mathop{\mbox{Tr}}\nolimits}
\nc{\Det}{\mathop{\mbox{Det}}\nolimits}
\rnc{\det}{\mathop{\mbox{det}}\nolimits}
\nc{\rk}{\mathop{\mbox{rk}}\nolimits}
\nc{\del}{\partial}
\nc{\diag}{\mathop{\mbox{diag}}\nolimits}
\nc{\ra}{\rightarrow}
\nc{\Ra}{\Rightarrow}
\nc{\LRa}{\Leftrightarrow}
\nc{\lra}{\leftrightarrow}
\nc{\ot}{\otimes}
\rnc{\ss}{\subset}
\nc{\nul}{\noindent\underline}
\nc{\non}{\nonumber\\}
\nc{\mat}[4]{\left(\begin{array}{cc}#1&#2\\#3&#4\end{array}\right)}
\rnc{\lg}{\frak{g}}
\nc{\G}[3]{\Gamma^{#1}_{\;{#2}{#3}}}
\nc{\nam}{\nabla_{\mu}}
\nc{\nan}{\nabla_{\nu}}
\nc{\dx}{\dot{x}}
\nc{\dxl}{\dot{x}^{\la}}
\nc{\dxm}{\dot{x}^{\mu}}
\nc{\dxn}{\dot{x}^{\nu}}
\nc{\ddx}{\ddot{x}}
\nc{\ddxm}{\ddot{x}^{\mu}}
\nc{\ddxn}{\ddot{x}^{\nu}}
\nc{\dxi}{\dot{\xi}}
\nc{\ddxi}{\ddot{\xi}}

\def\ci{\cite}

\def\la{\label}
\def \p {\phi}

\begin{document}

\rightline{SISSA/102/2003/EP}

\vfill

\begin{center}
{\Large\sc Penrose Limits and Spacetime Singularities}
\end{center}

\begin{center}
{\large
M.\ Blau${}^{a}$\footnote{e-mail: {\tt matthias.blau@unine.ch}},
M. Borunda${}^{b}$\footnote{e-mail: {\tt mborunda@he.sissa.it}},
M.\ O'Loughlin${}^{b}$\footnote{e-mail: {\tt loughlin@sissa.it}},
G.\ Papadopoulos${}^{c}$\footnote{e-mail: {\tt gpapas@mth.kcl.ac.uk}}
}
\end{center}

\centerline{\it ${}^a$ Institut de Physique, Universit\'e de Neuch\^atel,
Rue Breguet 1}
\centerline{\it CH-2000 Neuch\^atel, Switzerland}

\centerline{\it ${}^b$ S.I.S.S.A. Scuola Internazionale Superiore di Studi
Avanzati}
\centerline{\it Via Beirut 4, I-34014 Trieste, Italy}

\centerline{\it ${}^c$ Department of Mathematics,  King's College London}
 \centerline{\it London WC2R 2LS, U.K. }

\vskip -2.0 cm

\begin{center}
{\bf Abstract}
\end{center}
\vskip -0.1 cm

We give a covariant characterisation of the Penrose plane wave limit: 
the plane wave profile matrix $A(u)$ is the restriction of the null
geodesic deviation matrix (curvature tensor) of the original spacetime
metric to the null geodesic, evaluated in a comoving frame. We also
consider the Penrose limits of spacetime singularities and show that
for a large class of black hole, cosmological and null singularities
(of Szekeres-Iyer ``power-law type''), including those of the FRW and
Schwarzschild metrics, the result is a singular homogeneous plane wave
with profile $A(u)\sim u^{-2}$, the scale invariance of the latter
reflecting the power-law behaviour of the singularities.

\vfill

\newpage

\setcounter{footnote}{0}

\section{Introduction}

The Penrose limit construction \cite{penrose} associates to every
spacetime metric and choice of null geodesic in that spacetime a plane
wave metric
\be
ds^2 = 2dudv + A_{ab}(u)x^a x^b du^2 + d\vec{x}^2\;\;.
\ee
Here $A_{ab}(u)$ is the plane wave profile matrix and the computation of
the Penrose limit along a null geodesic $\gamma$ amounts to determining
the matrix $A_{ab}(u)$ from the metric of the original spacetime.

This has recently been used to show \cite{bfhp2} that the
Penrose limit of the $AdS_5\times S^5$ IIB superstring background is the
BHFP maximally supersymmetric plane wave \ci{bfhp1}. String theory in
this RR background is exactly solvable \cite{rrm,mt}, giving rise
to a novel explicit form of the AdS/CFT correspondence \cite{bmn}.
Following these developments, many Penrose limits have been computed
for various supergravity backgrounds and their applications have been
explored.

Despite these developments, the precise nature of the Penrose limit and
the extent to which it encodes generally covariant properties of the
original spacetime have remained somewhat elusive, also because the usual
definition and practical implementations of the Penrose limit indeed
require taking a limit and look rather non-covariant.

The primary purpose of this note is to provide a {\em completely covariant
characterisation and definition of the Penrose limit} wave profile matrix
$A_{ab}(u)$ which does not require taking any limit and which shows that
$A_{ab}(u)$ directly encodes diffeomorphism invariant information about
the original spacetime metric. Specifically, we will show that
\be
A_{ab}(u) = -R_{aubu}|_{\gamma(u)}\;\;,
\label{keyeq}
\ee
where $R$ is the curvature tensor of the original metric, and the
components refer to a parallel-transported frame along the null geodesic
with $ds^2= 2E^+E^- + \d_{ab}E^aE^b$ and $E_+=\del_u$. Thus $A_{ab}(u)$,
which is uniquely determined by these conditions up to constant orthogonal
transformations, is nothing other than the standard \cite[Section
4.2]{hawkingellis} {\em transverse frequency matrix} of the null geodesic
deviation equation of the original metric.

In particular, therefore, since lightcone gauge string theory on a plane
wave becomes a two-dimensional theory of free fields with mass (frequency)
matrix $(-A_{ab})$ \cite{rrm,mt}, we see that the components of the mass
matrix of the fields are certain components of the curvature tensor of the
spacetime before the Penrose limit is taken. Thus imaginary frequencies,
which lead to tachyonic worldsheet modes in the lightcone gauge
(but need not signal an instability \cite{bgs,dmlz}), correspond to
diverging null geodesics in the original spacetime.

Since singularities of $A_{ab}(u)$ result from curvature singularities
of the original spacetime, it is of interest to analyse the nature
of Penrose limits of spacetime singularities, as they encode information
about the rate of growth of curvature and geodesic deviation as one
approaches the singularity of the original spacetime along a null
geodesic. 

What we will demonstrate (referring to \cite{bbop2} for more details)
is that in this case the Penrose limits exhibit a 
remarkably universal behaviour in the sense that for a large class
of black hole, cosmological and null singularities, indeed all the
Szekeres-Iyer metrics \cite{SI,CS} with singularities of 
``power-law type'', one obtains plane wave metrics of the form
\be
ds^2 = 2 dudv + A_{ab}x^a x^b \frac{du^2}{u^2} + d\vec{x}^2\;\;,
\ee
with $A_{ab}$ constant and eigenvalues bounded by $1/4$.  Due to the
existence of the scale invariance $(u,v)\ra(\lambda u,\lambda^{-1}v)$,
these singular plane waves are {\em homogeneous} \cite{bfp,prt,hpw},
reflecting the scaling behaviour of the original power-law singularities.

String theory in singular homogeneous plane wave backgrounds is exactly
solvable \cite{sanchez,prt,mmga}. It has also been shown that in a
class of such models the string oscillator modes can be analytically
continued across the singularity \cite{prt}. Since the Penrose limit can
be considered as the origin of a string expansion around the original
background \cite{bfp} (see \cite{glmssw} for the AdS case), the above
observations about the relation of these backgrounds to interesting
spacetime singularities provide additional impetus for understanding
string theory in an expansion around such metrics.

\section{A Covariant Characterisation of the Penrose Limit}

\subsection{The Penrose Limit}

The Penrose limit \cite{penrose} associates to every spacetime metric
$g_{\mu\nu}$ and choice of null geodesic $\gamma$ in that spacetime a
(limiting) plane wave metric. The first step is to rewrite the metric
in coordinates adapted to $\gamma$, Penrose coordinates, as
\be             
ds^2 = 2dU dV + a(U,V,Y^k) dV^2 + 2b_i(U,V,Y^k) dV dY^i + 
g_{ij}(U,V,Y^k)dY^i dY^j
\label{pcs}
\ee        
This corresponds to an embedding of $\gamma$ into a twist-free congruence
of null geodesics, given by $V$ and $Y^k$ constant, with $U$ playing the role
of the affine parameter and $\gamma(U)$ coinciding with the geodesic at
$V=Y^k=0$. 

The next step is to perform the change of coordinates ($\Omega\in\RR$)
\be
(U,V,Y^{k})=(u,\Omega^2\bar{v},\Omega y_{k}) 
\;\;.
\ee
The Penrose limit metric $\bar{g}_{\mu\nu}$ is then defined by 
\be
d\bar{s}^2 = \lim_{\Omega\ra 0}\Omega^{-2}ds_{\Omega}^2
= 2dud\bar{v} + \bar{g}_{ij}(U)dy^i dy^j\;\;,
\label{rcs}
\ee
where $ds_{\Omega}^2$ is the  metric $ds^2$ in the coordinates
$(u,\bar{v},y^i)$ and $\bar{g}_{ij}(U)= g_{ij}(U,0,0)$. This is the metric
of a plane wave in Rosen coordinates. Pragmatically speaking, the Penrose
limit metric is obtained by setting the components $a$ and $b_i$ of the
metric to zero and restricting $g_{ij}$ to the null geodesic $\gamma$.

A coordinate transformation $(u,\bar{v},y^k)
\ra (u,v,x^a)$ puts the metric into the standard Brinkmann form
\be
d\bar{s}^2 = 2dudv + A_{ab}(u)x^ax^b du^2 + d\vec{x}^2\;\;.
\ee
Here
\be
A_{ab}(u) = -\bar{R}_{aubu}(u)
= -\bar{R}_{iuju}(u)\bar{E}^{i}_{a}(u)\bar{E}^{j}_{b}(u)\;\;,
\label{aree}
\ee
with $\bar{R}_{aubu}(u)$ the only non-vanishing component of the 
Riemann curvature tensor of $\bar{g}_{\mu\nu}$. $\bar{E}^{i}_{a}$
is an orthonormal coframe for the transverse metric $\bar{g}_{ij}$,
satisfying the symmetry condition
\be
\dot{\bar{E}}_{ai}\bar{E}^{i}_{b} =\dot{\bar{E}}_{bi}\bar{E}^{i}_{a}\;\;.
\label{sc1}
\ee

\subsection{Curvature and Penrose Limits}

We now establish the relation between the wave profile $A_{ab}(u)$
of the Penrose limit metric and certain components of the curvature
tensor of the original metric.

We consider the components $R^{i}_{\;UjU}$ of the curvature tensor of
the metric (\ref{pcs}) which enter into the geodesic deviation equation
of the corresponding null geodesic congruence. The first observation is
that
\be
R^{i}_{\;UjU} = -(\del_{U}\Gamma^{i}_{\;jU} +
\Gamma^{i}_{\;kU}\Gamma^{k}_{\;jU})
\ee
does not depend on the coefficients $a$ and $b_i$ of the metric and only
involves $U$-derivatives of $g_{ij}$. It follows that these components
of the curvature tensor are related to those of the Penrose limit metric
by
\be
\bar{R}^{i}_{\;uju} = R^{i}_{\;UjU}|_{\gamma} 
\label{rrbar}
\ee
Next we introduce a pseudo-orthonormal frame $E^{M}_{\mu}$,
$M=(+,-,a)$ for the metric (\ref{pcs}),
\be
ds^2 = 2 E^+ E^- + \d_{ab}E^a E^b\;\;,
\label{pof}
\ee
which is parallel along the null geodesic congruence,
$\nabla_{U}E^{M}_{\mu}=0$. We choose $E_+=\del_U$ to be
tangent to the geodesics. Then it is not difficult to see that
$E_a$ has the form
\be
E_a = E_a^i\del_{i} + E_{a}^{U}\del_U\;\;,
\label{eav}
\ee
where $E^{a}_{i}$ is a vielbein for $g_{ij}(U,V,Y^K)$ satisfying
\be
\dot{E}_{ai}E^{i}_{b} =\dot{E}_{bi}E^{i}_{a}\;\;.
\label{sc2}
\ee
This condition is independent of $a,b_i$ and only involves $U$-derivatives
of $E^{a}_{i}$. We can thus conclude that the vielbeins $\bar{E}^{a}_{i}$
of the Penrose limit metric satisfying the symmetry condition (\ref{sc1})
can be obtained from the parallel-propagated (\ref{sc2}) vielbeins of the 
full metric by restriction to the null geodesic $\gamma$,
\be
\bar{E}^{a}_{i} = E^{a}_{i}|_{\gamma}\;\;.
\label{eebar}
\ee
Combining (\ref{aree}) with (\ref{rrbar}) and (\ref{eebar}), and using
(\ref{eav}) we thus obtain the key result that the frequency matrix
(wave profile) $A_{ab}(u)$ of the Penrose limit metric is the transverse
null geodesic deviation matrix of the original metric, evaluated in a
parallel propagated frame \cite[Section 4.2]{hawkingellis},
\be
A_{ab}(u) = -(R_{i+j+}E^{i}_{a}E^{j}_{b})|_{\gamma}\;\;.
\ee
As a consequence, even though we had to appeal to Penrose adapted
coordinates (\ref{pcs}) to implement the standard definition 
(\ref{rcs}) of the
Penrose limit, we now arrive at a {\em fully covariant characterisation
and definition of the Penrose limit}. While this is implied by what we 
have already said, it may be worth reiterating it: 

Given a null geodesic $\gamma$, one constructs a pseudo-orthonormal
parallel propagated coframe $(E_+,E_-,E_a)$ with $E_+=\del_u$ tangent
to the null geodesic and $E_-$ characterised by $g(E_-,E_-)=0$
and $g(E_+,E_-)=1$. Then the Penrose limit is the plane wave metric
characterised by the wave profile
\be
A_{ab}(u) = - R_{a+b+}|_{\gamma}\;\;,
\ee
which is determined uniquely up to $u$-independent orthogonal 
transformations. 

\section{Singular Homogeneous Plane Waves from Penrose Limits of Spacetime
          Singularities}

Given the above results it is of interest to study Penrose limits
of spacetime singularities, as they encode diffeomorphism invariant
information about the rate of growth of curvature and geodesic deviation
along a null geodesic.

We will see that for a large class of spacelike, timelike or null
singularities the Penrose limit metric is that of a singular 
homogeneous plane wave \cite{prt,hpw}, 
\be
ds^2 = 2 dudv + A_{ab}x^a x^b \frac{du^2}{u^2} + d\vec{x}^2
\label{2}
\ee
exhibiting the scale invariance $(u,v)\ra(\lambda u,\lambda^{-1}v)$.
This has already been observed before in a variety of other special 
brane and cosmological backgrounds \ci{bfp,fis,patricot,kunze}.
Note that, because of the scale invariance, this metric is uniquely
characterised by the eigenvalues of $A_{ab}$ (up to permutations), in
contrast to the case of symmetric plane waves ($A_{ab}(u)$ constant)
for which the eigenvalues are scaled by $\lambda^2$ under the above
transformation.

\subsection{Examples: FRW and Schwarzschild Plane Waves}

We briefly illustrate this with some examples which will be dicussed in
more detail in \cite{bbop2}.

The $D=(n+1)$-dimensional FRW metric
\be
ds^2 = -dt^2 + a(t)^2(dr^2 + f_k(r)^2 d\Omega_{n-1}^2)\;\;,
\ee
($f_k(r)=r,\sin r,\sinh r$ for $k=0,+1,-1$ respectively)
has a unique Penrose limit, characterised by the diagonal wave profile
$A_{ab}=A(u)\d_{ab}$ with
\be
A(u) = -\frac{8\pi G}{n-1} \frac{\rho(u) + p(u)}{a(u)^2}\;\;.
\ee
Here $\rho$ and $p$ are the energy-density and pressure of the perfect fluid,
and $a(u)=a(t(u))$ etc., with $a(t)$ determined by the Friedmann equations
and $t(u)$ determined by the null geodesic condition with $u$ the affine 
parameter. Specialising to the equation of state $p=w\rho$, one finds that
near the singularity $A(u)$ behaves as\footnote{This is also the general
behaviour of the $k=0$ metrics and generalises the
result reported in \cite{bfp}.}
\be
A(u) = - \frac{h}{(1+h)^2} u^{-2}\;\;,
\label{Au}
\ee
where $h=2/n(1+w)$. Thus this is a singular homogeneous plane wave,
with frequency squares bounded by $1/4$ (which is attained for $h=1$).
This is interesting because it is known \cite{prt,hpw} that the behaviour
of particles and strings in the background (\ref{2}) is qualitatively
different for frequency squares less than or greater than $1/4$. 

For the $D=(d+2)\geq 4$-dimensional Schwarzschild metric
\be
ds^2 = -f(r)dt^2 + f(r)^{-1} dr^2 + r^2 d\Omega_d^2\;\;,
\label{ssm}
\ee
where $f(r) = 1 - 2m/r^{d-1}$, one finds that
$A_{ab}(u)$ is diagonal, with \cite{mbictplec} 
\be 
A_{22}(u)=\ldots = A_{dd}(u) = -\frac{(d+1)m\ell^2}{r(u)^{d+3}}
\ee
and (since this is a vacuum plane wave) $A_{11} = - \sum_{a=2}^{d}A_{aa}$.
Here $r(u)$ is the solution to the null geodesic equation and $\ell\neq 0$
the angular momentum. Thus the resulting
plane wave metric is singular iff the original null geodesic runs into the
singularity, which will happen for sufficiently small values of $\ell$. 
In this case, as $r(u)\ra 0$ one has
\be
A_{22}(u)=\ldots = A_{dd}(u) = - \frac{2(d+1)}{(d+3)^2} u^{-2}\;\;.
\ee
Again this is a singular homogeneous plane wave with frequencies bounded
by $1/4$. Note that the behaviour near the singularity depends only on
the spacetime dimension $D=d+2$, neither on the mass parameter $m$ of the
black hole nor on the angular momentum $\ell$ of the null geodesic used
to approach the singularity. Curiously, the frequency squares obtained
in the Penrose limit of the Schwarzschild metric are precisely those of
a dust-filled FRW universe of the same dimension (for example, $6/25$
for $D=4$).

Thus string and particle propagation in the near-singularity regime of
the Schwarzschild plane wave is identical to that in the $w=0$ FRW
plane wave in all but one of the transverse directions. The appearance
of an imaginary frequency in the remaining transverse direction is
of course dictated by the fact that the Schwarzschild plane wave is,
unlike the FRW plane wave, a vacuum solution.

\subsection{A general Result: 
Penrose Limits of Szekeres-Iyer Power-Law Type Singularities}

From a purely calculational point of view, the occurrence of $u^{-2}$-type
plane waves with frequency squares bounded by $1/4$ can be attributed to
the fact that in Rosen coordinates (\ref{rcs}) the leading behaviour in
$U$ of the transverse metric $\bar{g}_{ij}(U)$ close to the singularity
is of power-law type \cite{prt,hpw}, which is reflected in the scale
invariance of the Penrose limit. Thus to assess the generality of this
kind of result, one needs to enquire about the generality of spacetime
singularities exhibiting such a power-law behaviour. 

In \cite{SI} (see also \cite{CS}), in the context of investigations of
the Cosmic Censorship Hypothesis, Szekeres and Iyer studied a large
class of four-dimensional spherically symmetric metrics they dubbed
``metrics with power-law type singularities''. Such metrics encompass
all the FRW metrics, Lema\^itre-Tolman-Bondi dust solutions,
cosmological singularities of the Lifshitz-Khalatnikov type,
as well as other types of metrics with null singularities. 
In ``double-null form'', these metrics take the form 
\be
ds^2 = 2\ex{A(U,V)}dUdV + \ex{B(U,V)} d\Omega_2^2\;\;,
\label{sim}
\ee
where $A(U,V)$ and $B(U,V)$ have expansions
\bea
A(U,V)&=& p \ln X(U,V) + \mathrm{regular\; terms}\non
B(U,V)&=& q \ln X(U,V) + \mathrm{regular\; terms}
\eea
near the singularity surface $X(U,V)=0$. 
The residual coordinate transformations $U\ra U'(U)$, $V\ra V'(V)$
preserving the form of the metric (\ref{sim}) can be used to make
$X(U,V)$ linear in $U$ and $V$,
\be
X(U,V) = aU+bV\;\;,\;\;\;\;\;\;a,b=\pm 1, 0\;\;,
\ee
with $ab=\pm 1,0$ corresponding to timelike, spacelike and null singularities
respectively. This choice of gauge essentially fixes the coordinates 
uniquely, and thus the ``critical exponents'' $p$ and $q$ contain
diffeomorphism invariant information. The Schwarzschild metric, for
example, has $p=-1/2$ and $q=1$, as is readily seen by starting with 
the metric in Eddington-Finkelstein or Kruskal-Szekeres coordinates
and transforming to the Szekeres-Iyer gauge.

It is now a simple matter to determine all the Penrose limits of the
near-singularity metrics
\be
ds^2 = 2 X^p dUdV + X^q d\Omega_2^2\;\;,
\ee
in order to study the genericity of the $u^{-2}$-behaviour with frequency
squares bounded by $1/4$. One finds that for generic values of $p$
and $q$ within certain regions of the $(p,q)$-plane the behaviour as
$X(u)\ra 0$ is always of this type. For example, from this point of view
the $\pm 6/25$ of the Schwarzschild metric arise as $q(p+2)/(p+q+2)^2$
and $p(q+2)/(p+q+2)^2$ for $p=-1/2$, $q=1$.

Curiously, the resulting diagram in the $(p,q)$-plane delineating
the regions with $u^{-2}$-behaviour from other possibilities
bears a tantalising resemblance to the Szekeres-Iyer phase diagram
\cite[Fig.1]{SI} of near-singularity energy-momentum tensors.

\vspace{-0.3cm}
\subsection*{Acknowledgements}

This work has been supported by the European Community's Human Potential
Programme under contract HPRN-CT-2000-00131 Quantum Spacetime and by
the Swiss National Science Foundation.  MBo thanks CONACyT (Mexico) and
is grateful to the Institut de Physique, Neuch\^atel, for hospitality
during the final stages of this work.  GP thanks the Abdus Salam ICTP
for hospitality where part of this project was completed.

\vspace*{-0.3cm}
\rnc{\Large}{\normalsize}


\begin{thebibliography}{00}
\addcontentsline{toc}{section}{References}
\frenchspacing
\begin{small}
\addtolength{\itemsep}{-8pt}

\bibitem{penrose} R. Penrose, {\em Any space-time has a plane wave as a
limit}, in {\em Differential geometry and relativity}, Reidel, Dordrecht
(1976) pp.~271--275.

\bibitem{bfhp2} M. Blau, J. Figueroa-O'Farrill, C. Hull, G. Papadopoulos,
{\em Penrose  limits  and  maximal  supersymmetry}, Class. Quant. Grav. 19
(2002) L87-L95, {\tt hep-th/0201081}.

\bibitem{bfhp1} M. Blau, J. Figueroa-O'Farrill, C. Hull, G. Papadopoulos,
{\em A  new  maximally  supersymmetric background of IIB superstring
theory}, JHEP 0201 (2002) 047, {\tt hep-th/0110242}.

\bibitem{rrm} R.R.Metsaev, {\em Type  IIB  Green-Schwarz  superstring
in plane wave Ramond-Ramond background}, Nucl. Phys. B625 (2002) 70-96,
{\tt hep-th/0112044}.

\bibitem{mt} R.R. Metsaev, A.A. Tseytlin, {\em Exactly solvable model
of superstring in Ramond-Ramond plane wave background}, Phys.Rev. D65
(2002) 126004, {\tt hep-th/0202109}.

\bibitem{bmn} D. Berenstein, J. Maldacena, H. Nastase, {\em Strings  in
flat space and pp waves from ${\cal N}=4$ Super Yang Mills}, JHEP 0204
(2002) 013, {\tt hep-th/0202021}.

\bibitem{hawkingellis} S. Hawking, G. Ellis, {\em The Large Scale Structure
of Space-Time}, Cambridge University Press (1973).

\bibitem{bgs} D. Brecher, J. P. Gregory, P. M. Saffin, {\em String theory
and the Classical Stability of Plane Waves}, 
Phys. Rev. D67 (2003) 045014, {\tt hep-th/0210308}.

\bibitem{dmlz} D. Marolf, L. Pando Zayas, {\em On the Singularity
Structure and Stability of Plane Waves}, JHEP 0301 (2003) 076,
{\tt hep-th/0210309}.

\bibitem{bbop2} M. Blau, M. Borunda, M. O'Loughlin, G. Papadopoulos,
{\em in preparation}.

\bibitem{SI} P. Szekeres, V. Iyer, {\em Spherically Symmetric Singularities
and Strong Cosmic Censorship}, Phys. Rev. D 47 (1993) 4362-4371.

\bibitem{CS} M. C\'el\'erier, P. Szekeres, {\em Timelike and null focusing
singularities in spherical symmetry: a solution to the cosmological horizon 
problem and a challenge to the cosmic censorship hypothesis}, Phys.Rev. D65
(2002) 123516, {\tt gr-qc/0203094}.

\bibitem{bfp} M. Blau, J. Figueroa-O'Farrill, G. Papadopoulos, {\em
Penrose limits, supergravity and brane dynamics}, Class. Quant. Grav. 19
(2002) 4753, {\tt hep-th/0202111}.

\bibitem{prt} G. Papadopoulos, J.G. Russo and A.A. Tseytlin,
{\em Solvable model of strings in a time-dependent plane-wave background},
Class. Quant. Grav. 20 (2003) 969, {\tt hep-th/0211289}.

\bibitem{hpw} M. Blau, M. O'Loughlin, {\em Homogeneous Plane Waves},
Nucl. Phys. B654 (2003) 135-176, {\tt hep-th/0212135}.

\bibitem{sanchez} H.J. de Vega, N. Sanchez, {\em Strings falling into
spacetime singularities}, Phys. Rev. D45 (1992) 2783-2793; H.J. de
Vega, M. Ramon Medrano, N.  Sanchez, {\em Classical and quantum strings
near spacetime singularities: gravitational plane waves with arbitrary
polarization}, Class. Quant. Grav.  10 (1993) 2007-2019.

\bibitem{mmga} M. Blau, M. O'Loughlin, G. Papadopoulos, A. Tseytlin, {\em
Solvable  Models  of  Strings  in  Homogeneous  Plane Wave Backgrounds},
Nucl. Phys. B673  (2003) 57-97, {\tt hep-th/0304198}.

\bibitem{glmssw}
C. Callan, H. Lee, T. McLoughlin, J. Schwarz, I. Swanson, X. Wu,
{\em Quantizing String Theory in $AdS_5 \times S^5$: Beyond the
pp-Wave}, Nucl. Phys. B673 (2003) 3-40, {\tt hep-th/0307032}.

\bibitem{fis} H. Fuji, K. Ito, Y. Sekino, {\em Penrose Limit and
String Theories on Various Brane Backgrounds}, JHEP 0211 (2002) 005,
{\tt hep-th/0209004}.

\bibitem{patricot} C. Patricot, {\em Kaigorodov spaces and their Penrose
limits},  Class. Quant. Grav. 20 (2003) 2087-2102, {\tt  hep-th/0302073}.

\bibitem{kunze} K. Kunze, {\em T-Duality and Penrose limits of spatially
homogeneous and inhomogeneous cosmologies}, Phys. Rev. D68 (2003) 063517,
{\tt gr-qc/0303038}.

\bibitem{mbictplec} M. Blau, {\em Plane Waves and Penrose Limits},
Lecture Notes for the ICTP School on Mathematics in String and Field Theory 
(June 2-13 2003).

\end{small}
\end{thebibliography}
\end{document}